\documentstyle[11pt,newpasp,twoside,epsf,graphics,natbib]{article}

\bibpunct{(}{)}{;}{a}{}{,}

\def\chandra{{\it Chandra }}
\def\XMM{{\it XMM-Newton }}
\def\xmm{{\it XMM }}
\def\iras13349{IRAS~13349+2438}
\def\mcg6{MCG$-$6-30-15}
\def\grs1915{GRS~1915+105}

\def\edcomment#1{\iffalse\marginpar{\raggedright\sl#1\/}\else\relax\fi}
\reversemarginpar

\begin{document}
\title{Probing X-ray Emitting Plasma with High Resolution Chandra and XMM-Newton Spectra}
\vspace{-0.2in}
 \author{Julia C. Lee}
\affil{Chandra Fellow,  MIT Center for Space Research, 77 Massachusetts Ave. 
	NE80-6073, Cambridge, MA. 02139 USA}

\vspace{-0.1in}
\begin{abstract}
Highlights of interesting astrophysical discoveries are reviewed in
the context of high resolution X-ray spectroscopy  made possible with
\chandra and \XMM, and its relevance to  atomic physics calculations
and measurements is discussed.   These spectra have shown
that the overlap between astrophysics and atomic physics is
stronger than ever, as discoveries of new X-ray lines and edge
structure is matching the need for increasingly detailed theoretical
calculations and experimental measurements of atomic data.
\end{abstract}

\vspace{-0.4in}
\section{X-ray spectral probes of astrophysical systems at high resolution}
High resolution X-ray spectroscopy provides a powerful new tool for
advancing our understanding of the physical environments of  energetic
astrophysical systems.  As demonstrated with \chandra and \xmm
spectral studies, the scientific impact is far reaching, encompassing
studies of stars, supernova remnants (SNR),  X-ray binaries (XRBs),
active galactic nuclei (AGN),  clusters, and the interstellar and
intergalactic medium (respectively ISM and IGM).  To give a flavor for
some  of the newly X-ray discovered spectral features and their
relevance to spectral modeling and calculations, I will draw mostly on
examples from observations of AGN and  XRBs with which  I have been
involved.  See \cite{highres_review:03} for a complete review of
\chandra and \xmm results.

%\vspace{-0.1in}
\subsection{Narrow emission and absorption lines}
%\vspace{-0.05in}
Narrow (i.e. barely resolved) emission and absorption lines are nearly
ubiquitous in the  astrophysical sources seen at high
resolution.  From the lines strengths alone, we can deduce much about
the conditions of the plasma, which range from the ``X-ray cold''
where fluorescent emission and photoionzation prevail to
the ``X-ray hot'' where collisional ionization dominates.  From the view of
atomic calculations and spectral modeling, the parameterization of the
emitters and absorbers are at an advanced state as demonstrated
by high resolution spectral studies of the photoionized plasma in
Seyfert galaxies \citep[e.g.,][]{sako_mrk3:00,  ogle_ngc4151:00,
br_mcg6:01, ngc4051:01, jcl_mcg6_wa1:01, jcl_mcg6conf:02,
kaspi3783:00,kaspi3783:01,
kaspi3783:02,bngc1068:02,kngc1068:02,blustin_ngc7469:03,
sako_mcg6:03}.  Specific features which demonstrate the power of
high resolution spectroscopy come from the detection of high order
(low oscillator strength) resonance absorption lines (i.e. higher than
Lyman~$\gamma$) which are the mark of high optical depth clouds
\citep[e.g.,][]{jcl_mcg6_wa1:01, kaspi3783:02}.
Commonly used atomic codes include
\footnote{http://www.nublado.org/}\texttt{Cloudy:}~\citet{cloudy_ref}, 
\footnote{http://heasarc.gsfc.nasa.gov/docs/software/xstar/xstar.html}\texttt{XSTAR:}~\citet{xstar_ref} and
\footnote{http://xmm.astro.columbia.edu/research.html}\texttt{photoion:}~\citet{photoion_ref}.

Strong evidence for non-equilibrium collisionally ionized plasma can be seen in SNRs 
observed with \chandra and \xmm.  One of the best examples is 1E~0102.2-7219 where
spectral-line images reveal progressive ionizations in the remnant 
attributed to a reverse shock (\citealt{kaf_eo102:03}; and 
Fig.~5 of \citealt{kafspieEO102}).

\subsection{P-Cygni Profiles, Line Variability, Doppler Shifts, and Spatial-Spectral Doppler mapping}
Outflows have been seen in many different forms in \chandra and  \xmm
spectra.  Key spectroscopic signatures include  (1) Doppler-shifted
absorption and/or emission lines which provide information on the
kinematics and geometry of the outflow,  (2) P-Cygni profiles
(red-shifted/rest-frame emission from material out of the
line-of-sight, accompanied by blue-shifted absorption lines from the
foreground, line-of-sight part of the wind) as seen in both X-ray
binaries \citep[e.g. Circinus~X1:][]{brandt_cirx1:00,
schulz_cirx1:02}  and AGN \citep[e.g. NGC~3783:][]{kaspi3783:01,
kaspi3783:02}, and (3)  more subtle  variability effects
\citep[e.g. the micro-quasar GRS~1915+105:][]{jcl_grs1915:02}.  Of
these, the most remarkable are those which show relativistic
velocities  (e.g. the BHC SS~433 where $v_{\rm jet} \sim 0.27c$,
\citealt{hlm_ss433:02}),  and more recently also seen in QSOs and Narrow
line Seyfert galaxies.   From the X-ray measurements of these lines
and shifts, a great deal can be learned about the X-ray portion of the
flow. For example, based on the line broadening, information
about the flow opening angle can be deduced while density diagnostics
using observed He-like lines can provide important limits on  the mass
flow rate.  For some of the brighter SNRs, spatial-spectral Doppler
mapping can be used to reveal the 3-D structure of the SNR 
(e.g. 1E~0102, Fig~6 of \citealt{kafspieEO102}) -- see \citet{dewey_snr} for technique;
for Cassiopeia~A, see \citet{casA:2002}.

%For some of the brighter SNRs, spatial-spectral Doppler
%mapping can be used to determine the location and conditions of the
%shock regions.

\subsection{Inner Shell lines}
`Low' ionization (in the X-ray sense) lines such as O~{\sc iii-vi}  or
Fe~{\sc vi-xvii} are typically only seen in other wave-bands
(e.g. UV), but photoexcitation of the ion's inner shell electron
followed by  auto-ionization causes resonance lines to appear in the
X-ray band.  These lines have been detected  in the X-ray spectra of
AGN : (1) the broad structure between $\sim 15.5 - 17$~\AA\, ($\sim
0.72-0.8$~keV) known as the `unresolved transition array' (UTA) of
inner-shell 2p-3d resonance absorption lines in weakly ionized M-shell
Fe~{\sc vi-xii}  \citep[for calculations, see][]{feuta_calc:01} was
first detected in the Seyfert galaxy \iras13349 \citep{sako13349:01}
and since seen in NGC~3783 \citep{kaspi3783:02},  (2) similarly, the
K-L resonance absorption (inner shell 1s-2p transition) of  Li-like
oxygen \citep[for calculations, see][]{oxygen6calc:00} was also first
discovered in a Seyfert galaxy  \citep[\mcg6:][]{jcl_mcg6_wa1:01}.
Since then, lower ionization oxygen lines have also been seen in this
source (\citealt{sako_mcg6:03}; Lee et al. in prep.), and inner shell lines of
Si~{\sc vii-xii} and S~{\sc xii-xiv} have been reported in
NGC~3783~\citep{kaspi3783:02}.  Low ionization oxygen lines have also
been  detected in the X-ray binaries and attributed to the
line-of-sight ISM  or that intrinsic to the source
\citep[e.g.,][]{xrb_paerels:01,ism_takei:02, juett_ism:03} .

For highly extincted extragalactic sources where a UV spectrum cannot
be seen, the inner shell lines can eventually provide a powerful
alternative for studying the ``lukewarm'' part of the partially
ionized gas in the AGN environment (i.e. the warm absorber).  For ISM
studies, these lines provide an important diagnostic  for the
abundance distributions in our local Universe.

\subsection{Edge Structure (XAFS) and Shifts}
\vspace{-0.05in}
Photoelectric edges are seen as prominent spectral features in  X-ray
spectra of sources with significant absorption, from  which we can
deduce the optical depth and hydrogen column of the  absorbing medium.
However, based on the edge discontinuity alone, we cannot distinguish
between gas versus dust phase absorption.   In some cases, dust
has been inferred as the source of an Fe~L photoelectric edge
at $\sim 17.7$~\AA\, ($\sim 0.7$~keV) 
\citep[e.g.,][]{jcl_mcg6_wa1:01, jcl_mcg6conf:02}.
However, the most direct probe of dust is if the X-ray Absorption
Fine Structure (XAFS), which probes material in solid form can be extracted
from high resolution data of bright highly absorbed XRBs.  Tentative 
detections of these features have been reported in the \chandra 
spectrum of \grs1915 \citep{jcl_grs1915:02}.

\vspace{-0.15in}
\section{Concluding Thoughts}
\vspace{-0.10in}
The wealth of high resolution \chandra and \XMM spectra  accumulated
over the course of the last $\sim$ four years  has provided us with a
very rich laboratory for probing the details of astrophysical plasma.
While the atomic physics calculations appear to be well suited to
model the high resolution X-ray data in hand, much of the detailed
modeling has been parametric.  Our next step should be to take
the wealth of atomic and satellite data and connect it to the detailed
models of the astrophysical sources themselves.

\noindent\acknowledgements \small I thank Claude Canizares for his generosity \&
insightful discussions.  I also thank the MIT CXC \& HETGS members.
I am grateful for the financial support of the
Chandra fellowship grant PF2--30023,  issued by the
Chandra X-ray Observatory Center which is operated by SAO for
NASA under contract NAS8--39073.

\vspace{-0.1in}
%%%%%%%

%%%%%%%%%%%%%%%%%%%%%%%%%%%%%%%%%%%%%%%%%%%%%%%%%%%%%%%%%%%%%%%%%%%%%%%%%%%%%%%%%


\begin{thebibliography}{}
\small
\bibitem[\protect\astroncite{{Behar}, {Sako} \& {Kahn}}{2001}]{feuta_calc:01}
{Behar}, E., {Sako}, M., \& {Kahn}, S.~M.,  2001, \apj, 563, 497

\bibitem[\protect\astroncite{{Blustin} et~al.}{2003}]{blustin_ngc7469:03}
{Blustin}, A.~J., et~al., 2003, \aap, 403, 481

\bibitem[\protect\astroncite{{Brandt} \& {Schulz}}{2000}]{brandt_cirx1:00}
{Brandt}, W.~N., \& {Schulz}, N.~S.,  2000, \apjl, 544, L123

\bibitem[\protect\astroncite{{Branduardi-Raymont} et~al.}{2001}]{br_mcg6:01}
{Branduardi-Raymont}, G., {Sako}, M., {Kahn}, S.~M., {Brinkman}, A.~C.,
  {Kaastra}, J.~S., \& {Page}, M.~J.,  2001, \aap, 365, L140

\bibitem[\protect\astroncite{{Brinkman} et~al.}{2002}]{bngc1068:02}
{Brinkman}, A.~C., {Kaastra}, J.~S., {van der Meer}, R.~L.~J., {Kinkhabwala},
  A., {Behar}, E., {Kahn}, S.~M., {Paerels}, F.~B.~S., \& {Sako}, M.,  2002,
  \aap, 396, 761

\bibitem[\protect\astroncite{{Collinge} et~al.}{2001}]{ngc4051:01}
{Collinge}, M.~J., et~al., 2001, \apj, 557, 2

\bibitem[\protect\astroncite{{Dewey}}{2002}]{dewey_snr}
{Dewey}, D.,  2002,
\newblock in High Resolution X-ray Spectroscopy with XMM-Newton and Chandra

\bibitem[\protect\astroncite{{Ferland} et~al.}{1998}]{cloudy_ref}
{Ferland}, G.~J., {Korista}, K.~T., {Verner}, D.~A., {Ferguson}, J.~W.,
  {Kingdon}, J.~B., \& {Verner}, E.~M.,  1998, \pasp, 110, 761

\bibitem[\protect\astroncite{{Flanagan} et~al.}{2003a}]{kafspieEO102}
%{Flanagan}, K.~A., {Canizares}, C.~R., {Dewey}, D., {Fredericks}, A., {Houck},
%  J.~C., {Lee}, J.~C., {Marshall}, H.~L., \& {Schattenburg}, M.~L.,  2003a,
{Flanagan}, K.~A., et al., 2003a,
\newblock in X-Ray and Gamma-Ray Telescopes and Instruments for Astronomy.
  Edited by Joachim E. Truemper, Harvey D. Tananbaum. Proceedings of the SPIE,
  Volume 4851, pp. 45-56 (2003)., 45

\bibitem[\protect\astroncite{{Flanagan} et~al.}{2003b}]{kaf_eo102:03}
{Flanagan}, K.~A., {Canizares}, C.~R., {Dewey}, D., {Houck}, J.~C.,
  {Fredericks}, A.~C., {Schattenburg}, M.~L., {Markert}, T., \& {Davis}, D.,
  2003b, \apj, submitted

\bibitem[\protect\astroncite{{Juett}, {Schulz} \&
  {Chakrabarty}}{2003}]{juett_ism:03}
{Juett}, A.~M., {Schulz}, N.~S., \& {Chakrabarty}, D.,  2003, \apj, submitted

\bibitem[\protect\astroncite{{Kallman} \& {Bautista}}{2001}]{xstar_ref}
{Kallman}, T., \& {Bautista}, M.,  2001, \apjs, 133, 221

\bibitem[\protect\astroncite{{Kaspi} et~al.}{2002}]{kaspi3783:02}
{Kaspi}, S., et~al., 2002, \apj, 574, 643

\bibitem[\protect\astroncite{{Kaspi} et~al.}{2001}]{kaspi3783:01}
{Kaspi}, S., et~al., 2001, \apj, 554, 216

\bibitem[\protect\astroncite{{Kaspi} et~al.}{2000}]{kaspi3783:00}
{Kaspi}, S., {Brandt}, W.~N., {Netzer}, H., {Sambruna}, R., {Chartas}, G.,
  {Garmire}, G.~P., \& {Nousek}, J.~A.,  2000, \apjl, 535, L17

\bibitem[\protect\astroncite{{Kinkhabwala} et~al.}{2003}]{photoion_ref}
{Kinkhabwala}, A., {Behar}, E., {Sako}, M., {Gu}, M.~F., {Kahn}, S.~M., \&
  {Paerels}, F.,  2003, \apj,  submitted

\bibitem[\protect\astroncite{{Kinkhabwala} et~al.}{2002}]{kngc1068:02}
{Kinkhabwala}, A., et~al., 2002, \apj, 575, 732

\bibitem[\protect\astroncite{{Lee} et~al.}{2002a}]{jcl_mcg6conf:02}
{Lee}, J.~C., {Canizares}, C.~R., {Fang}, T., {Morales}, R., {Fabian}, A.~C.,
  {Marshall}, H.~L., \& {Schulz}, N.~S.,  2002a,
\newblock in X-ray Spectroscopy of AGN with Chandra and XMM, 9

\bibitem[\protect\astroncite{{Lee} et~al.}{}]{jcl_mcg6_wa2}
{Lee}, J.~C., {Fang}, T., {Kallman}, T., {Canizares}, C.~R., {Fabian}, A., R.,
  G.~R., \& {Marshall}, H.~L., \apj, in preparation

\bibitem[\protect\astroncite{{Lee} et~al.}{2001}]{jcl_mcg6_wa1:01}
{Lee}, J.~C., {Ogle}, P.~M., {Canizares}, C.~R., {Marshall}, H.~L., {Schulz},
  N.~S., {Morales}, R., {Fabian}, A.~C., \& {Iwasawa}, K.,  2001, \apjl, 554,
  L13
\bibitem[\protect\astroncite{{Lee} et~al.}{2002b}]{jcl_grs1915:02}
{Lee}, J.~C., {Reynolds}, C.~S., {Remillard}, R., {Schulz}, N.~S., {Blackman},
  E.~G., \& {Fabian}, A.~C.,  2002b, \apj, 567, 1102

\bibitem[\protect\astroncite{{Marshall}, {Canizares} \&
  {Schulz}}{2002}]{hlm_ss433:02}
{Marshall}, H.~L., {Canizares}, C.~R., \& {Schulz}, N.~S.,  2002, \apj, 564,
  941

\bibitem[\protect\astroncite{{Ogle} et~al.}{2000}]{ogle_ngc4151:00}
{Ogle}, P.~M., {Marshall}, H.~L., {Lee}, J.~C., \& {Canizares}, C.~R.,  2000,
  \apjl, 545, L81

\bibitem[\protect\astroncite{{Paerels} et~al.}{2001}]{xrb_paerels:01}
{Paerels}, F., et~al., 2001, \apj, 546, 338

\bibitem[\protect\astroncite{Paerels \& Kahn}{2003}]{highres_review:03}
Paerels, F.~B., \& Kahn, S.~M.,  2003, Annual Review of Astronomy and
  Astrophysics, 41, 291

\bibitem[\protect\astroncite{{Pradhan}}{2000}]{oxygen6calc:00}
{Pradhan}, A.~K.,  2000, \apjl, 545, L165

\bibitem[\protect\astroncite{{Sako} et~al.}{2001}]{sako13349:01}
{Sako}, M., et~al., 2001, \aap, 365, L168

\bibitem[\protect\astroncite{{Sako} et~al.}{2003}]{sako_mcg6:03}
{Sako}, M., et~al., 2003, \apj,  in press

\bibitem[\protect\astroncite{{Sako} et~al.}{2000}]{sako_mrk3:00}
{Sako}, M., {Kahn}, S.~M., {Paerels}, F., \& {Liedahl}, D.~A.,  2000, \apjl,
  543, L115

\bibitem[\protect\astroncite{{Schulz} \& {Brandt}}{2002}]{schulz_cirx1:02}
{Schulz}, N.~S., \& {Brandt}, W.~N.,  2002, \apj, 572, 971

\bibitem[\protect\astroncite{{Takei} et~al.}{2002}]{ism_takei:02}
{Takei}, Y., {Fujimoto}, R., {Mitsuda}, K., \& {Onaka}, T.,  2002, \apj, 581,
  307

\bibitem[\protect\astroncite{{Willingale} et~al.}{2002}]{casA:2002}
{Willingale}, R., {Bleeker}, J.~A.~M., {van der Heyden}, K.~J., {Kaastra},
  J.~S., \& {Vink}, J.,  2002, \aap, 381, 1039


%\bibliographystyle{jwapjbib}
%\bibliography{jcl}
\end{thebibliography}
\end{document}